\documentclass[showpacs,aps,graphicx]{revtex4}%,twocolumn
\usepackage{amsmath}%preprint,
\usepackage{graphicx}

\usepackage{subfigure}

\begin{document}

\title{Entanglement distillation  for quantum communication network with atomic-ensemble memories\footnote{Published in Opt. Express \textbf{22}, 23897 (2014)}}

\author{Tao Li$^1$,  Guo-Jian Yang$^1$,  and Fu-Guo Deng$^{1,2,}\footnote{Corresponding author: fgdeng@bnu.edu.cn}$}

\address{$^1$Department of Physics, Applied Optics Beijing Area Major Laboratory,
Beijing Normal University, Beijing 100875, China\\
$^2$State key Laboratory of Networking and Switching Technology,
Beijing University of Posts and Telecommunications, Beijing 100876,
China}

%\email{*fgdeng@bnu.edu.cn}

\date{\today }

\begin{abstract}
Atomic ensembles are effective memory nodes for quantum
communication network  due to the  long coherence time and the
collective enhancement effect for the nonlinear interaction between
an ensemble and a  photon. Here we investigate the possibility of
achieving the entanglement distillation for nonlocal atomic
ensembles by the input-output process of a single photon as a result
of cavity quantum electrodynamics.  We give an optimal entanglement
concentration protocol (ECP) for two-atomic-ensemble systems in a
partially entangled pure state with known parameters and  an
efficient ECP for the systems in an unknown partially entangled pure
state with a nondestructive parity-check detector (PCD). For the
systems in a mixed entangled state, we introduce an entanglement
purification protocol with PCDs. These entanglement distillation
protocols have high fidelity and efficiency with current
experimental techniques, and they are useful for quantum
communication network  with atomic-ensemble memories.
\end{abstract}
 \maketitle

%\pacs{03.67.Pp Quantum error correction and other methods for
%protection against decoherence - 03.67.Hk Quantum communication}

%\ocis{(270.0270) Quantum optics; (060.5565)   Quantum communications; (270.5580) Quantum electrodynamics; (020.0020)  Atomic and molecular physics.}

 % REPLACE WITH CORRECT OCIS CODES FOR YOUR ARTICLE

\section{Introduction}

%Some quantum communication protocols work with
%maximally entangled quantum systems
%\cite{QKD1,QKD2,QKD3,QSS1,QSDC1,QSDC2}.

%
%,QSS2,QSS3

Quantum entanglement plays an important role in quantum
communication, such as quantum teleportation   \cite{Teleportation},
quantum dense coding (QDC) \cite{DC1,DC2}, quantum key distribution
 \cite{Qnoise2002,QKD1,QKD2,QKD3}, quantum secret sharing
 \cite{QSS1}, quantum secure direct communication \cite{QSDC1,QSDC2},   and so
on.  Quantum teleportation    \cite{Teleportation} requires a
maximally entangled photon pair to set up the quantum channel for
teleporting an unknown single-particle quantum state without moving
the particle itself.   The high capacity of QDC \cite{DC1,DC2} comes
from the maximal entanglement of quantum systems. However, an
entangled photon pair is usually produced locally and suffers
inevitably from  the environment noise (e.g., the thermal
fluctuation, vibration, imperfection of an optical fiber, and
birefringence effects \cite{lixhapl}) in its distribution process
between the two parties in quantum communication, which will degrade
its entanglement or even make it in a mixed state. The process of
quantum state storage will degrade the entanglement as well. This
decoherence will decrease the security of quantum communication
protocols and the fidelity of quantum teleportation.  In a practical
long-distance quantum communication network, a quantum repeater
\cite{Qrepeater1998} is required to depress the decoherence by the
destructive effect of the noise\cite{Qrepeater1999dur}. Entanglement
distillation
\cite{Bennett1996,Deutsch,Pan1,Simon,Pan2,shengprA2008,RenEPPLPL,
DEPPtwostep,DEPPsheng,DEPPli,DEPPdeng,DEPPShengLPL,Bennett1996ECP,Bose1999swap,Zhao2001ECP,
Yamamoto2001ECP,Sheng2008ECP,Sheng2012ECP,Ren2013ECP,RenHECP2014,reviewECP}
is one of the key techniques in a quantum repeater protocol
\cite{Qrepeater1998,Qrepeater1999dur} and it is used to distil
highly entangled states from less entangled ones
\cite{Bennett1996,Deutsch,Pan1}.  It includes entanglement
purification and entanglement concentration.

%Entanglement purification
%can be used for faithful teleportation via noisy channels \cite{Bennett1996} and quantum privacy
%amplification in quantum cryptography \cite{Deutsch}.

% and improved its security.

%
%, but not its efficiency.

Accurately, entanglement purification is used to obtain a subset of
high-fidelity nonlocal entangled quantum systems from a set of those
in a mixed state with less entanglement
\cite{Bennett1996,Deutsch,Pan1,Simon,Pan2,shengprA2008,RenEPPLPL,DEPPtwostep,DEPPsheng,DEPPli}.
In 1996, Bennett \emph{et al}. \cite{Bennett1996} proposed the first
entanglement purification protocol (EPP) to purify a particular
Werner state, resorting to quantum controlled-not (CNOT) gates. In
2001, Pan \emph{et al}.  \cite{Pan1} introduced an EPP with linear
optical elements,  by keeping the cases that the two photons owned
by the same party have the identical polarization. In 2002, Simon
and Pan \cite{Simon} developed an EPP with linear optical elements
and a currently available parametric down-conversion (PDC) source.
Recently, Ren and Deng presented an EPP for spatial-polarization
hyperentangled photon systems \cite{RenEPPLPL} and Sheng \emph{et
al.} proposed some deterministic EPPs
\cite{DEPPtwostep,DEPPsheng,DEPPli,DEPPdeng,DEPPShengLPL}.
Entanglement concentration is used to get some nonlocal maximally
entangled systems from a set of systems in a partially entangled
pure state.  In 1996, Bennett \emph{et al.}  \cite{Bennett1996ECP}
presented  the first entanglement concentration protocol (ECP) for
two-particle systems, resorting to the Schmidt projection.  In 2001,
Zhao \emph{et al.}
 \cite{Zhao2001ECP} and  Yamamoto \emph{et al.}  \cite{Yamamoto2001ECP}
proposed two similar ECPs independently with linear optical
elements.  In 2008, Sheng \emph{et al.}  \cite{Sheng2008ECP}
developed an ECP with cross-Kerr nonlinearity and its efficiency was
improved largely by iteration of the entanglement concentration
process.  In 2013, Ren, Du, and Deng  \cite{Ren2013ECP} presented an
ECP for hyperentangled photon pairs with the parameter-splitting
method
 based on linear optical elements. In 2014, Ren and Long
\cite{RenHECP2014} gave a general scheme for polarization-spatial
hyperentanglement concentration of photon pairs. By far, some ECPs
for atom systems \cite{atomECP2,atomECP3,atomECP4} have been
proposed.

%a seminal paper by Duan\emph{et al.}

The atomic ensemble system is one of the most promising candidates
for quantum communication  \cite{RMP2011QR}, due to the collective
enhancement effect originating from the indistinguishability of the
atoms interacting with radiations. Moreover, the time for the
storage of a single photon in a cold atom ensemble is about several
milliseconds \cite{Qmemory2009} and atom ensembles can, in
principle,  be used as memory nodes for long-distance quantum
communication network. For example, in \cite{DLCZ}, an  atomic
ensemble is used as a local memory node \cite{Qmemory2009} for
global quantum communication. Inspired by the repeater protocol in
\cite{DLCZ}, a number of  works for long-distance quantum
communication based on atomic ensembles have been done
\cite{RMP2011QR}. In 2007, Chen \emph{et al.} \cite{ZhaoChenQR}
introduced a fault-tolerant quantum repeater protocol with atomic
ensembles and linear optics. In 2010, Zhao \emph{et al.}
\cite{Zhao2010EPP} proposed a quantum repeater protocol for the
atomic ensembles by utilizing the Rydberg blockade effect
\cite{RMP2010Rydberg1}. In 2011, Aghamalyan \emph{et al.}
\cite{QRsinglephoton2011} proposed a quantum repeater protocol based
on the deterministic storage of a single photon in an atomic
ensemble with the cavity-assisted interaction.

In the fault-tolerant quantum repeater \cite{ZhaoChenQR}, two atomic
ensembles that associate with the photons in different polarizations
are used  as an effective memory node. The entanglement purification
for the nonlocal atomic ensembles can be performed with the EPP for
polarizing photons \cite{Pan1} when the quantum memories are
involved \cite{Qmemory2009}. In \cite{Zhao2010EPP}, two different
single collective spin-wave excitation states of the atomic ensemble
are used to encode a stationary qubit. The entanglement purification
for the atomic ensembles in two different nodes is completed with
the scheme proposed in \cite{Bennett1996}, since the CNOT gate
between the two atomic ensembles in each node is available, when
they are placed within the blockade radius \cite{RMP2010Rydberg1}.

In this paper, we investigate the possibility of achieving the
entanglement distillation for nonlocal atomic ensembles trapped in
single-sided small cavities
\cite{QRsinglephoton2011,MeiQIC2009,RMP2013Cavityensemble,singleside1,singleside2}
by the input-output process of single photons. We give three
efficient entanglement distillation protocols (EDPs) for nonlocal
two-atomic-ensemble systems, including  an optimal  ECP for a
partially entangled state with known parameters,  an ECP for  an
unknown partially entangled pure state with a nondestructive
parity-check detector (PCD) which works efficiently in a simple way,
and an EPP for  a mixed entangled state. Compared with the EDPs for
atom systems by others \cite{atomECP2,atomECP3,atomECP4}, our EDPs
for nonlocal atomic ensembles have the advantage of  high fidelity
and efficiency with current experimental techniques. Moreover, it is
easier to implement our EDPs than the former and they are useful for
quantum communication network with atomic ensembles acting as
memories.

%\begin{verbatim}7.3
\begin{figure}[htbp]             %Figure 1
\centering\includegraphics[width=10 cm]{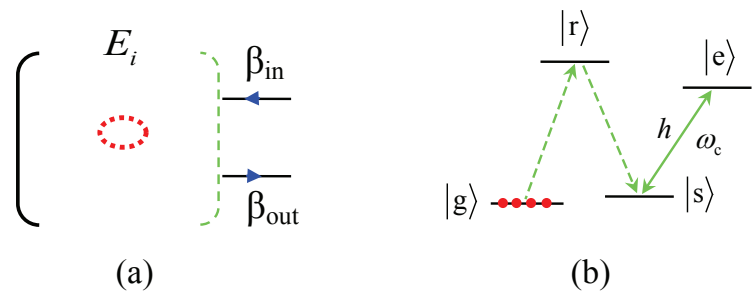} \caption{ (a)
Schematic diagram for  a single-sided cavity coupled to an atomic
ensemble system composed of $N$ cold atoms. (b)  Schematic diagram
for the level structure of a cold atom. }\label{fig1}
\end{figure}

\section{Nondestructive parity-check detector on two local atomic ensembles}

\subsection{An atomic-ensemble-cavity system}

Let us  consider an ensemble composed of $N$ cold atoms trapped in a
single-sided optical cavity
\cite{QRsinglephoton2011,MeiQIC2009,RMP2013Cavityensemble,singleside1,singleside2},
in which one of the mirrors is perfectly reflective and another one
is of small transmission allowing for incoupling and outcoupling to
light, shown in Fig. \ref{fig1}(a). The atom has a four-level
internal structure, shown in Fig. \ref{fig1}(b).
 The two hyperfine ground states of a cold atom are denoted with $|g\rangle$ and
$|s\rangle$. The excited state $|e\rangle$ and the Rydberg state
$|r\rangle$ are two auxiliary states. The atomic transition between
$|s\rangle$ and $|e\rangle$ is resonantly coupled to the polarized
cavity mode $a$, which is nearly resonantly driven by the input
photon in the polarization $|h\rangle$ with the frequency $\omega$,
while the transition between $|g\rangle$ and $|e\rangle$ is a
dipole-forbidden one
 \cite{Duan2004QIP}. Initially, the atomic ensemble is prepared in the
ground state $|G\rangle=|g_1,\dots,g_{_N}\rangle$. An arbitrary
unitary operation between the ground state $|G\rangle$ and the
single collective spin-wave excitation state
$|S\rangle=\frac{1}{\sqrt{N}}\sum_{j}|g_1,\dots,s_j,\dots,g_{_N}\rangle$
can be performed efficiently with the Rydberg blockade effect of the
state $|r\rangle$ by the collective laser manipulations on the
ensemble  \cite{EnsembleQIP,RMP2010Rydberg1}. When the Rydberg
blockade shift is $\Delta=2\pi\times100MHz$, the transition between
$|G\rangle$ and $|S\rangle$ can be completed with an effective
coupling strength $\Omega=2\pi\times1MHz$ and the probability of
nonexcited and doubly excited errors are about $10^{-3}-10^{-4}$
\cite{Saffman,MeiQIC2009}.

In the frame rotating with the  cavity frequency $\omega_c$, the
Hamiltonian of the whole system composed of an input photon and an
atom ensemble inside a single-sided cavity  can be expressed as
($\hbar=1$)  \cite{Duan2004QIP,quantumoptics}
\begin{eqnarray}                  %%%%%%%%%%%%%%%%%%%%%%%%%%%%%%%%Equation 1
\begin{split}
H
&\;\;=\;\sum_{j=1}^{N}\left[\left(\Delta-i\frac{\gamma_{e_{j}}}{2}\right)\sigma_{e_{j}{e_{j}}}
+ig_j\left(a\sigma_{e_{j}{s_{j}}}-a^+\sigma_{s_{j}{e_{j}}}\right)\right] \\
&\;\;\;\;\;\;\;+i\sqrt{\frac{\kappa}{2\pi}}\int{}d\delta'\left[b^+(\delta')a-b(\delta')a^+\right]
+\int d\delta'b^+(\delta')b(\delta').
\end{split} \label{Hamiltonian}
\end{eqnarray}
Here $a$ and $b$ are the respective annihilation operators of the
$|h\rangle$ polarized cavity mode and the input photon mode with
$[a,a^+]=1$ and
$[b(\delta'),b^+(\delta'')]=\delta(\delta'-\delta'')$ (hereafter
$\delta$ refers to Dirac delta function, $\delta'$ and $\delta''$
represent the detunings). $\Delta=\omega_0-\omega_c$ is the detuning
between the cavity mode frequency $\omega_c$ and the dipole
transition frequency $\omega_0$. $\delta'= \omega -\omega_c$,
$\sigma_{e_{j}{e_{j}}}=|e_{j}\rangle\langle e_{j}|$, and
$\sigma_{e_{j}{s_{j}}}=|e_{j}\rangle\langle s_{j}|$. The coupling
strength $\sqrt{\frac{\kappa}{2\pi}}$ between the cavity and the
input photon is taken to be a real constant
 \cite{Duan2004QIP,quantumoptics}, since only the fields with the carrier
frequency close to $\omega_c$ contribute mostly to the cavity mode.
The coefficients $\gamma_{e_{j}}$ and $g_{j}$ denote the spontaneous
emission rate of the excited state $|e_{j}\rangle$ and the coupling
strength  between the \emph{j-th} atom and the cavity mode $a$,
respectively. For simplicity, we assume $g_j=g$ and
$\gamma_{e_{j}}=\gamma$ below.

Initially, when the ensemble is prepared in the state $|S\rangle$,
the input photon is in the state $|h\rangle$, and the cavity mode is
vacuum, with the Hamiltonian in Eq. (\ref{Hamiltonian}), the
evolution of the whole system can be confined in the first-order
excitation subspace in  a general state $|\Psi(t)\rangle$. Here
\begin{eqnarray}                %%%%%%%%%%%%%%%%%%%%%%%%%%%%%%%%Equation 2
|\Psi(t)\rangle =\alpha(t)|S\rangle \otimes|1,0\rangle +
\int{}d\delta'\beta(\delta',t)|S\rangle \otimes |0,1\rangle
+\zeta|E\rangle\otimes|0,0\rangle, \label{gstate}
\end{eqnarray}
where
$|E\rangle=\frac{1}{\sqrt{N}}\sum_{j}|g_1,\dots,e_j,\dots,g_{_N}\rangle$.
$|m,n\rangle$ represents the Fock state with the photon numbers $m$
(0 or 1) and $n$ (0 or 1) in the cavity mode and the free space
mode, respectively. The Schr\"{o}dinger equations for this system
can be specified as:
\begin{eqnarray}\label{sequa}   %%%%%%%%%%%%%%%%%%%%%%%%%%%%%Equation 3-5
i\dot{\alpha}(t) &=&
-ig\zeta(t)-i\sqrt{\frac{\kappa}{2\pi}}\int{}d\delta'\beta(\delta',t),\label{se1}
\\ i\dot{\beta}(\delta',t)&=& i\sqrt{\frac{\kappa}{2\pi}}\alpha(t)+\delta'\beta(\delta',t),\label{se2}\\
i\dot{\zeta}_j(t) &=&
(\Delta-i\frac{\gamma}{2})\zeta_j(t)+ig\alpha(t).\label{se3}
\end{eqnarray}
In the condition  $ t_0 < t < t_1$ (the times $t_0$ and $t_1$
correspond to the moments that the pulse goes in and  comes out of
the cavity with the presumable pulse shapes given by
$\beta^{in}(t_0)$ and $\beta^{out}(t_1)$,  respectively
\cite{quantumoptics}), one can get the standard input-output
relation from Eq. (\ref{se2}), i.e.,
\begin{eqnarray}          %%%%%%%%%%%%%%%%%%%%%%%%%%%%%%%%Equation 6
\beta^{out}(t) =
\beta^{in}(t)+\sqrt{\kappa}\,\alpha(t),\label{inputoutputt}
\end{eqnarray}
where
\begin{eqnarray}          %%%%%%%%%%%%%%%%%%%%%%%%%%%%%%%%Equation7
\beta^{in}(t)\!\!= \!\! \frac{1}{\sqrt{2\pi}}\int
e^{-i\delta'{}(t-t_0)}\beta(\delta',t_0)d\delta',\;\;\;\;\;\;\;\;\;
\beta^{out}(t)\!\!=\!\!\frac{1}{\sqrt{2\pi}}\int
e^{-i\delta'{}(t-t_1)}\beta (\delta',t_1)d\delta'.
\end{eqnarray}
Here $\beta^{in}(t)$ and $\beta^{out}(t)$ can be interpreted as the
input and the output to the single-sided cavity.
$\beta(\delta',t_0)$ and $\beta(\delta',t_1)$ are the probability
amplitudes of the input photon  with the frequency
$\omega=\omega_c+\delta'$ at the times $t_0$ and $t_1$,
respectively. Taking Eqs. (\ref{se1}) and (\ref{se3}) into account,
one can get the reflection coefficient
$r{(\delta')}=\beta(\delta',t_1)/\beta(\delta',t_0)$ of the cavity,
that is,
\begin{eqnarray}                  %%%%%%%%%%%%%%%%%%%%%%%%%%%%%%% Equation 8
r{(\delta')}=\frac{(\delta'-i\kappa/2)(\Delta+i\gamma/2)-g^2}{(\delta'+i\kappa/2)(\Delta+i\gamma/2)-g^2}.
\label{r}
\end{eqnarray}

If the ensemble is initially in the state $|G\rangle$, it will be
decoupled to the cavity mode and the  input photon in the
polarization $|h\rangle$ will be reflected by an empty cavity
\cite{quantumoptics}. The reflection coefficient $r_{_0}(\delta')$
is
\begin{eqnarray}              %%%%%%%%%%%%%%%%%%%%%%%%%%%%%%% Equation 9
r_{_0}{(\delta')}=\frac{\delta'-i\kappa/2}{\delta'+i\kappa/2}.
\label{r0}
\end{eqnarray}
Note that when the detuning $|\delta'|\ll \kappa$ and
$\gamma\kappa/4\ll{}g^2$,  one can get a unit reflection with
$r_{_0}{(\delta')}\simeq-1$ and $r{(\delta')}\simeq1$, respectively,
which can be summarized to an ideal reflection operator $\hat{R}$
when the input-output process is involved. Here
\begin{eqnarray}              %%%%%%%%%%%%%%%%%%%%%%%%%%%%%%% Equation 9'
\hat{R}=|h\rangle\langle{}h|(-|G\rangle\langle{}G|+|S\rangle\langle{}S|).
\label{Rideal}
\end{eqnarray}

\subsection{PCD on two local atomic ensembles with the input-output process of a single photon}

\begin{figure}[!h]%[tpb]                                              %Figure1 4(Color online)
\begin{center}
\includegraphics[width=9 cm,angle=0]{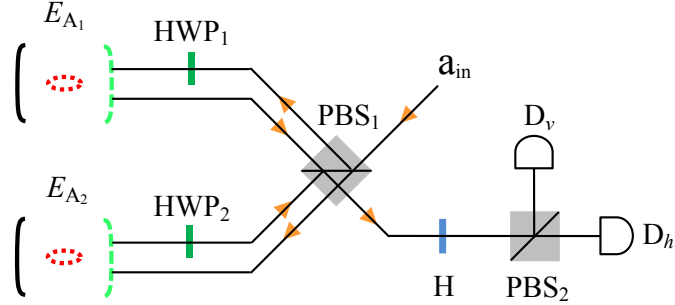}
\caption{ Schematic diagram for a PCD on two ensembles $E_{A_1}$ and
$E_{A_2}$. $a_{_{in}}$ is the input port of the photon.  HWP$_i$
($i=1,2$) represents a half-wave plate whose optical axis is set to
$\pi/4$ to perform the bit-flip operation
$\sigma_x=|h\rangle\langle{}v|+|v\rangle\langle{}h|$ on the photon.
$H$ represents a half-wave plate whose optical axis is set to
$\pi/8$ and completes the Hadamard transformation. PBS is a
polarizing beam splitter, which transmits the $|h\rangle$
polarization photon and reflects the $|v\rangle$ polarization
photon, respectively.}\label{fig2}
\end{center}
\end{figure}

The principle of our  nondestructive  PCD  on two local atomic
ensembles, say $E_{A_1}$ and $E_{A_2}$, is shown in Fig. \ref{fig2}.
Suppose the ensembles $E_{A_1}E_{A_2}$ are in the state
$|\varphi\rangle_{E_{A_1}E_{A_2}}=
\alpha_1|GG\rangle+\alpha_2|GS\rangle+\alpha_3|SG\rangle+\alpha_4|SS\rangle$.
A probe photon \emph{a} in the polarization state
$|\phi\rangle=\frac{1}{\sqrt{2}}(|h\rangle+|v\rangle)$  sent into
the port $a_{in}$ is split into two components $|h\rangle$ and
$|v\rangle$ by the polarizing beam splitter $PBS_1$ and then is led
into the two cavities which contain the ensemble $E_{A_1}$ and the
ensemble $E_{A_2}$, respectively. The state of the system composed
of the photon \emph{a} and the ensembles $E_{A_1}E_{A_2}$ evolves as
follows:
\begin{eqnarray}     %%%%%%%%       %%%%%%%%%%%%%%%%%%%%%%% Equation10
\begin{split}
|\varphi\rangle_{p_{E_{A_1}E_{A_2}}} &=
|\phi\rangle|\varphi\rangle_{_{E_{A_1}E_{A_2}}}
\xrightarrow[HPW_1]{PBS_1}
\sigma_x^{A_1}|\varphi\rangle_{p_{E_{A_1}E_{A_2}}}
\xrightarrow[]{\hat{R}}
\hat{R}_{A_2}\hat{R}_{A_1}\sigma_x^{A_1}|\varphi\rangle_{p_{E_{A_1}E_{A_2}}}
\\ &\xrightarrow[PBS_1]{HPW_2} \sigma_x^{A_2}
\hat{R}_{A_2}\hat{R}_{A_1}\sigma_x^{A_1}|\varphi\rangle_{p_{E_{A_1}E_{A_2}}}.
\end{split}
\label{phipab}
\end{eqnarray}
Here $\sigma_x^i=|h\rangle\langle v|+ |v\rangle\langle h|$
represents  the bit-flip operator on the photon sent to  ensemble
$i$ ($i= A_1$ or $A_2$).
$\hat{R}_i=|h\rangle_i\langle{}h|(-|G\rangle\langle{}G|+|S\rangle\langle{}S|)_i$
is the reflection operator on  ensemble $i$ and the photon.
Subsequently, the photon \emph{a} is subjected to a Hadamard
transformation $H_p$
$\big[$$|h\rangle\leftrightarrow1/\sqrt{2}(|h\rangle+|v\rangle)$ and
$|v\rangle\leftrightarrow1/\sqrt{2}(|h\rangle-|v\rangle)$$\big]$
denoted by $H$ in Fig. \ref{fig2}, and the state of the whole system
evolves into
\begin{eqnarray}     %%%%%%%%%%%%%%%%%%%%%%%%%%%%%%% Equation11
|\varphi\rangle'_{p_{E_{A_1}E_{A_2}}}\!\!\!= \!
|h\rangle_p(\alpha_1|GG\rangle-\alpha_4|SS\rangle)_{A_1A_2} +
|v\rangle_p(\alpha_2|GS\rangle - \alpha_3|SG\rangle)_{A_1A_2}.
\label{PCD}
\end{eqnarray}
When an $|h\rangle$ photon is detected,  the two-ensemble system
$E_{A_1} E_{A_2}$ is in the even-parity state $\vert
\varphi\rangle_e=\frac{1}{\sqrt{|\alpha_1|^2+|\alpha_4|^2}}(\alpha_1|GG\rangle-\alpha_4|SS\rangle)_{A_1A_2}$.
In contrast, if a $|v\rangle$ photon is detected, $E_{A_1} E_{A_2}$
is in the odd-parity state
$\vert\varphi'\rangle_o=\frac{1}{\sqrt{|\alpha_2|^2+|\alpha_3|^2}}(\alpha_2|GS\rangle
- \alpha_3|SG\rangle)_{A_1A_2}$. That is, the quantum circuit shown
in Fig. \ref{fig2} can be used to accomplish a PCD on the two atomic
ensembles $E_{A_1}E_{A_2}$. Moreover, it is quite simple and only
one efficient input-output process of a single photon  is involved.
It can be performed efficiently with a high fidelity since the
cavities show the unit reflectance.

\section{Entanglement concentration for partially entangled atomic ensembles}

\subsection{Optimal ECP for two nonlocal atomic ensembles in a known partially entangled state}

\begin{figure}[!h]%[tpb]                                              %Figure2 4(Color online)
\begin{center}
\includegraphics[width=13 cm,angle=0]{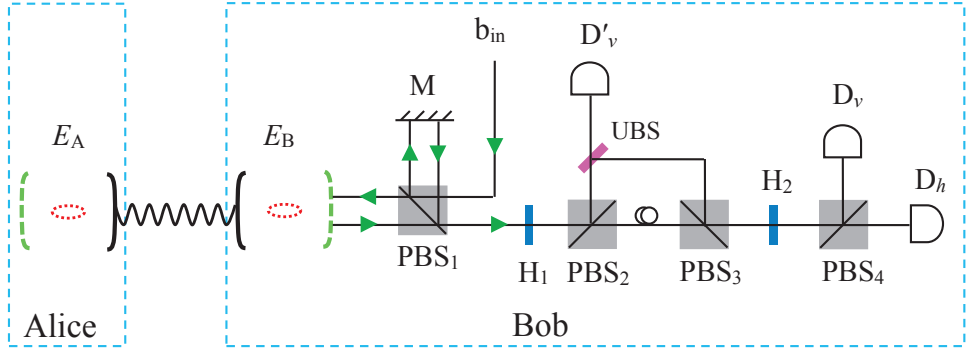}
\caption{Schematic diagram of our optimal ECP for a nonlocal
two-atomic-ensemble system in a partially entangled state with known
parameters.  Alice and Bob are two parties in two nonlocal memory
nodes in a quantum communication network. $E_A$ and $E_B$ are the
two nonlocal atomic ensembles which belong to Alice and Bob,
respectively.  The UBS is an unbalanced beam splitter with the
reflection coefficient $R=\alpha/\beta$.}\label{fig3}
\end{center}
\end{figure}

The principle of our optimal ECP  for a pair of nonlocal partially
entangled two-atomic-ensemble system $E_AE_B$ is shown in Fig.
\ref{fig3}. $E_A$ and $E_B$ belong to two parties in two nonlocal
memory nodes in a quantum communication network, say Alice and Bob,
respectively. This ECP is used to distill probabilistically a
nonlocal two-ensemble system in a maximally entangled Bell state
from that in a partially entangled pure state with known parameters.

Suppose the  system $E_AE_B$ is initially in the following partially
entangled  state
 \cite{Bose1999swap,Sheng2012ECP}
\begin{eqnarray}     %%%%%%%%%%%%%%%%%%%%%%%%%%%%%%%%Equation12
|\phi\rangle_{E_AE_B}=\alpha|GS\rangle_{AB}+\beta|SG\rangle_{AB},
\label{initialstateOECP}
\end{eqnarray}
where the  known  real coefficients $\alpha$ and $\beta$ satisfy
$|\alpha|^2 + |\beta|^2=1$ and $|\alpha|<|\beta|$ (the case of
$|\alpha|>|\beta|$ can be treated in the same way). In order to
complete this  optimal ECP, Bob prepares a single photon $b$ in the
state $|\phi\rangle_{b}=\frac{1}{\sqrt{2}}(|h\rangle+|v\rangle)$ and
leads it to the port $b_{in}$, shown in Fig. \ref{fig3}.

After PBS$_1$, the photon $b$ will be reflected by the cavity or the
mirror $M$, and a reflection operator
$\hat{R}=|h\rangle\langle{}h|(-|G\rangle\langle{}G|+|S\rangle\langle{}S|)$
is introduced. Subsequently, Bob performs a Hadamard operation on
the photon \emph{b} with the half-wave plate H$_1$.  The state of
the composite system composed of the photon \emph{b} and the
ensembles $E_A E_B$ evolves into $|\phi\rangle_{ABb}$,
\begin{eqnarray}      %%%%%%%%%%%%%%%%%%%%%%%%%%%%%%%%Equation13
|\phi\rangle_{ABb}=\alpha|GS\rangle_{AB}|h\rangle_{b} -
\beta|SG\rangle_{AB}|v\rangle_{b}.
\end{eqnarray}
PBS$_2$, PBS$_3$, and the unbalanced beam splitter (UBS)
\cite{Ren2013ECP} on the vertical path of the photon \emph{b} with
the reflection coefficient $R=\alpha/\beta$ will change
$|\phi\rangle_{ABb}$  into
\begin{eqnarray}     %%%%%%%%%%%%%%%%%%%%%%%%%%%%%%%%Equation14
|\phi'\rangle_{ABb}=\alpha(|GS\rangle_{AB}|h\rangle_{b}-|SG\rangle_{AB}|v\rangle_{b})-\sqrt{\beta^2-\alpha^2}|SG\rangle_{AB}|v_e\rangle_{b}.
\end{eqnarray}
The system is in the state $|SG\rangle_{AB}|v_e\rangle_{b}$ with the
probability $p_e=\beta^2-\alpha^2$, which leads to an error that can
be heralded by the click of the detector $D'_v$. If the detector
$D'_v$ does not click, the photon \emph{b} and the ensembles $E_A
E_B$ will be projected into the hybrid maximally entangled GHZ state
$|\phi''\rangle_{ABb}=\frac{1}{\sqrt{2}}(|GS\rangle_{AB}|h\rangle_{b}
- |SG\rangle_{AB}|v\rangle_{b})$, which takes place with the
probability  $2|\alpha|^2$.

To complete the  nonlocal optimal ECP for the entangled ensembles
$E_A E_B$, Bob performs another Hadamard operation on the photon
\emph{b} with H$_2$ and the state of the system becomes
\begin{eqnarray}      %%%%%%%%%%%%%%%%%%%%%%%%%%%%%%%%Equation15
|\phi''\rangle_{ABb}\rightarrow|\phi'''\rangle_{ABb}=\frac{1}{\sqrt{2}}(|h\rangle_b\otimes|\psi^{-}\rangle_{E_AE_B}
+|v\rangle_b\otimes|\psi^{+}\rangle_{E_AE_B}),
\end{eqnarray}
where
$|\psi^{\pm}\rangle_{E_AE_B}=\frac{1}{\sqrt{2}}(|GS\rangle\pm|SG\rangle)_{AB}$.
If the detector $D_v$ clicks, this optimal ECP is completed
successfully, and Alice and Bob share a nonlocal two-atomic-ensemble
system in a maximally entangled Bell state
$|\psi^{+}\rangle_{E_AE_B}$.  It is also completed successfully by
the response of the detector $D_h$, followed by a phase-flip
operation $\sigma_z=|G\rangle\langle{}G|-|S\rangle\langle{}S|$ on
the ensemble $E_B$. In other words, Bob can judge whether this ECP
succeeds or not, according to the response of the detectors. The
success probability of our optimal ECP is
$\eta^i_{c}=1-p_e=2|\alpha|^2$ which is just the maximal value in
nonlocal entanglement concentration for each nonlocal system.

Our optimal ECP can be generalized to distill maximally entangled
$N$-ensemble GHZ states from a partially entangled GHZ-class pure
state. Let us, for example, consider the case that Alice ($E_A$),
Bob ($E_B$), $\cdots{}$, and Charlie ($E_C$)  share a partially
entangled $N$-ensemble state
\begin{eqnarray}      %%%%%%%%%%%%%%%%%%%%%%%%%%%%%%%%Equation16
|\phi\rangle_{E_AE_B\cdots{}E_C}=\alpha|GG\cdots{}G\rangle_{AB\cdots{}C}+\beta|SS\cdots{}S\rangle_{AB\cdots{}C},
\end{eqnarray}
with $|\alpha|<|\beta|$ and $|\alpha|^2+|\beta|^2=1$. Through a
similar process to that for the two-ensemble case, the parties can
entangle a polarization photon with the $N$-ensemble system. A local
filtering operation on the $|v\rangle$ component of the photon
followed by a single-photon detection will project the remaining
N-ensemble system $E_A E_B \cdots E_C$ into the maximally entangled
GHZ state
%\begin{eqnarray}   %%%%%%%%%%%%%%%%%%%%%%%%%%%%%%%%Equation17
$|\phi'\rangle_{E_AE_B\cdots{}E_C}=\frac{1}{\sqrt{2}}
(|GG\cdots{}G\rangle_{AB\cdots{}C}+|SS\cdots{}S\rangle_{AB\cdots{}C})$
%\end{eqnarray}
with the  probability $P'_{mc}=2|\alpha|^2$.

\begin{figure}[!h]%[tpb]                                              %Figure2 4(Color online)
\begin{center}
\includegraphics[width=11 cm,angle=0]{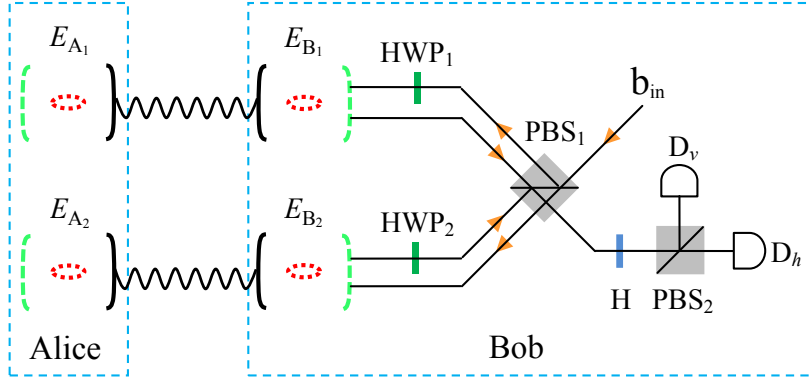}
\caption{ Schematic diagram of our ECP for a nonlocal
two-atomic-ensemble system in a partially entangled state with
unknown parameters, achieved by the input-output process of a single
photon. Bob completes the parity-check measurement on the ensembles
$E_{B 1}$ and   $E_{B 2}$ with a PCD, assisted by a single
photon.}\label{fig4}
\end{center}
\end{figure}

\subsection{ECP for  atomic ensembles in a partially entangled  state with unknown parameters}

The principle of our  ECP for  nonlocal atomic ensembles in a
partially entangled pure state with unknown parameters is shown in
Fig. \ref{fig4}. The two atomic ensembles $E_{A_1}$ and $E_{A_2}$
belong to Alice, and the two atomic ensembles $E_{B_1}$ and
$E_{B_2}$ belong to Bob. Suppose these two identical pairs of
partially entangled ensembles $E_{A_1}E_{B_1}$ and $E_{A_2}E_{B_2}$
are respectively in the states
\cite{Bennett1996ECP,Zhao2001ECP,Yamamoto2001ECP,Sheng2008ECP}
\begin{eqnarray}    %%%%%%%%%%%%%%%%%%%%%%%%%%%%%%%%Equation18
%\begin{split}
|\varphi\rangle_{E_{A_1}E_{B_1}}  = \alpha|GS\rangle_{ A_1 B_1 }+\beta|SG\rangle_{ A_1 B_1 },%\\
\;\;\;\;\;\;\;\;|\varphi\rangle_{E_{A_2}E_{B_2}}=
\alpha|GS\rangle_{ A_2 B_2 }+\beta|SG\rangle_{ A_2 B_2 },
%\end{split}
\label{two-ensemble}
\end{eqnarray}
where the coefficients $\alpha$ and $\beta$  are unknown and
$|\alpha|^2+|\beta|^2=1$. The essential process of the ECP is
projecting the state of $E_{B_1} E_{B_2}$ into the odd-parity
subspace other than the even-parity subspace after a bit-flip
operation on one of the photons in the ECPs
\cite{Zhao2001ECP,Yamamoto2001ECP}, and it succeeds in a heralded
way when the  detector $D_v$ clicks. Alice and Bob can distil a
maximally entangled two-ensemble system even in the case that a
practical input-output process of the single photon is involved,
since the even-parity state of $E_{B_1}E_{B_2}$ will never lead to
the click of $D_v$.

After Bob performs a parity-check measurement on his two atomic
ensembles $E_{B_1}$ and $E_{B_2}$,  if the photon detector $D_v$
clicks (an odd-parity outcome is obtained), the state of the
four-atomic-ensemble system $E_{A_1}E_{B_1}E_{A_2}E_{B_2}$ is
projected into a maximally entangled four-qubit GHZ state $
|\Phi\rangle_{E_{A_1}E_{B_1}E_{A_2}E_{B_2}}=\frac{1}{\sqrt{2}}(|GSSG\rangle-|SGGS\rangle)_{A_1B_1A_2B_2}$,
which takes place with the probability  $2|\alpha\beta|^2$. By
taking a Hadamard operation $H_E$ $\big[|G\rangle \leftrightarrow
\frac{1}{\sqrt{2}}(|G\rangle + |S\rangle), |S\rangle
\leftrightarrow\frac{1}{\sqrt{2}}(|G\rangle - |S\rangle)\big]$ on
both the ensembles $E_{A_2}$ and $E_{B_2}$, Alice and Bob evolve the
four-atomic-ensemble system into
\begin{eqnarray}        %%%%%%%%%%%%%%%%%%%%%%%%%%%%%%%%Equation 19
\begin{split}
|\Phi'\rangle_{E_{A_1}E_{B_1}E_{A_2}E_{B_2}}&\;\;=\;
\frac{1}{2\sqrt{2}}
\big[(|GS\rangle-|SG\rangle)_{A_1B_1}(|GG\rangle-|SS\rangle)_{A_2B_2}
\\&\;\;\;\;\;\;\;
+(|GS\rangle+|SG\rangle)_{A_1B_1}(|GS\rangle-|SG\rangle)_{A_2B_2}\big].
\end{split}
\end{eqnarray}
They can obtain a nonlocal two-atomic-ensemble system
$E_{A_1}E_{B_1}$ in a maximally entangled Bell state by measuring
the two ensembles $E_{A_2}$ and $E_{B_2}$ independently with the
basis $\{|G\rangle,|S\rangle\}$. In detail, if Alice and Bob obtain
two different outcomes ($|G\rangle_{A_2} |S\rangle_{B_2}$ or
$|S\rangle_{A_2}|G\rangle_{B_2}$), the remaining two-atomic-ensemble
system $E_{A_1}E_{B_1}$ will be projected into $
|\psi^+\rangle_{E_{A_1}E_{B_1}}$. If Alice and Bob obtain the same
outcomes $|G\rangle_{A_2} |G\rangle_{B_2}$ or $|S\rangle_{A_2}
|S\rangle_{B_2}$, the two-atomic-ensemble system $E_{A_1}E_{B_1}$ is
projected into the state $|\psi^-\rangle_{E_{A_1}E_{B_1}}$ which can
be transformed into the state $|\psi^+\rangle_{E_{A_1}E_{B_1}}$ with
a phase-flip operation
$\sigma_z=|G\rangle\langle{}G|-|S\rangle\langle{}S|$ on the ensemble
$E_{A_1}$ or $E_{B_1}$. In other words, they can get
$E_{A_1}E_{B_1}$ in the state $|\psi^+\rangle_{E_{A_1}E_{B_1}}$ with
the success probability $\eta^{i}_{c'}=2|\alpha\beta|^2$.

If the outcome of the parity-check measurement on $E_{B_1} E_{B_2}$
is an even-parity one (the $D_h$ detector clicks), the
four-atomic-ensemble system is projected into the partially
entangled state
$|\xi\rangle_{E_{A_1}E_{B_1}E_{A_2}E_{B_2}}=\frac{1}{\sqrt{|\alpha|^4+|\beta|^4}}(\alpha^2|GSGS\rangle
+ \beta^2|SGSG\rangle)_{A_1B_1A_2B_2}$. Alice and Bob can  measure
the ensembles $E_{A_2}$ and  $E_{B_2}$ with the basis
$\{|G\rangle,|S\rangle\}$ after a Hadamard operation $H_{E}$ on both
the two ensembles $E_{A_2}$ and  $E_{B_2}$, and they will obtain the
two-atomic-ensemble systems $E_{A_1}E_{B_1}$ in the state $|\varphi'
\rangle_{E_{A_1}E_{B_1}}=\frac{1}{\sqrt{|\alpha|^4+|\beta|^4}}(\alpha^2|GS\rangle
+ \beta^2|SG\rangle)_{A_1B_1}$ with or without a phase-flip
operation $\sigma_z$ on $E_{A_1}$. They can perform the ECP in the
second round by replacing $\alpha$ and $\beta$ with $\alpha'\equiv
\frac{\alpha^2}{\sqrt{|\alpha|^4+|\beta|^4}}$ and $\beta'\equiv
\frac{\beta^2}{\sqrt{|\alpha|^4+|\beta|^4}}$, respectively, when two
copies of two-atomic-ensemble systems in the state
$|\varphi'\rangle_{E_{A_1}E_{B_1}}$ are available. It is not
difficult to calculate the success probability of the ECP in the
second round
$\eta^{ii}_{c'}=\frac{2|\alpha\beta|^4}{(|\alpha|^4+|\beta|^4)^2}$.
After two rounds of entanglement concentration, the total success
probability is
$\eta^{i}_{c'_t}=\eta^{i}_{c'}+(\frac{1-\eta^{i}_{c'}}{2})\eta^{ii}_{c'}$.
Certainly,  the method described above can be cascaded to the $n$-th
round of concentration in the ideal case that the photon loss of the
input-output process and the decoherence of the ensembles are
negligible, and the total success probability of the ECP can be
further improved.

\section{ Entanglement purification for atomic ensemble systems with  PCDs}

%
%will be
%the process of entanglement

In general, a quantum system in a maximally entangled Bell  state
degrads into a mixed entangled state in its distribution between two
memory nodes and its storage. In this time, the parties  should use
an EPP to improve the entanglement of their nonlocal quantum systems
for creating a high-fidelity quantum channel in a quantum
communication network.

Suppose that the two-ensemble systems shared by Alice and Bob are in
a mixed entangled state $\rho_{E_AE_B}$
\cite{Pan1,Simon,Pan2,shengprA2008},
\begin{eqnarray}     %%%%%%%%%%%%%%%%%%%%%%%%%%%%%%%%Equation 20
\rho_{E_AE_B}=f_0|\psi^+\rangle\langle\psi^+|+(1-f_0)|\phi^+\rangle\langle\phi^+|.
\end{eqnarray}
It can be viewed as the mixture of two pure states
$|\psi^+\rangle=\frac{1}{\sqrt{2}}(|GS\rangle+|SG\rangle)_{AB}$ and
$|\phi^+\rangle=\frac{1}{\sqrt{2}}(|GG\rangle+|SS\rangle)_{AB}$ with
the probabilities $f_0$ and $1-f_0$, respectively. Here, we only
discuss the purification of the systems with the bit-flip error
$|\phi^+\rangle$, as the same as those for photon systems in
\cite{Pan1,Simon,Pan2,shengprA2008}, because the parties can convert
the phase-flip error into the bit-flip error with a Hadamard
operation $H_E$ on each of the two ensembles $E_A$ and $E_B$.

\begin{figure}[!h]%[tpb]                                              %Figure3 4(Color online)
\begin{center}
\includegraphics[width=13 cm,angle=0]{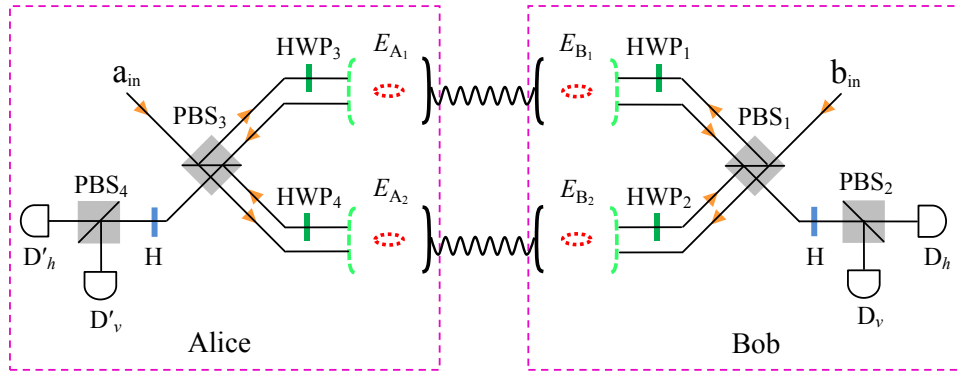}
\caption{ Schematic diagram of our EPP for atomic ensembles with
PCDs.}\label{fig5}
\end{center}
\end{figure}

The principle of our EPP for two-atomic-ensemble systems with PCDs
is shown in Fig. \ref{fig5}.  Alice and Bob choose two pairs of
two-ensemble systems $\rho_{E_{A_1}E_{B_1}}$ and
$\rho_{E_{A_2}E_{B_2}}$ in each time for purification. The
four-ensemble system $E_{A_1}E_{B_1}E_{A_2}E_{B_2}$ can be viewed as
the mixture of four pure states
$|\Psi_4\rangle_1=|\psi^+\rangle_{E_{A_1}E_{B_1}}\otimes|\psi^+\rangle_{E_{A_2}E_{B_2}}$,
$|\Psi_4\rangle_2=|\phi^+\rangle_{E_{A_1}E_{B_1}}\otimes|\psi^+\rangle_{E_{A_2}E_{B_2}}$,
$|\Psi_4\rangle_3=|\psi^+\rangle_{E_{A_1}E_{B_1}}\otimes|\phi^+\rangle_{E_{A_2}E_{B_2}}$,
and
$|\Psi_4\rangle_4=|\phi^+\rangle_{E_{A_1}E_{B_1}}\otimes|\phi^+\rangle_{E_{A_2}E_{B_2}}$
with the probabilities $f_0^2$, $f_0(1-f_0)$, $f_0(1-f_0)$, and
$(1-f_0)^2$, respectively. Alice prepares a single photon $a$ in the
state $|\varphi\rangle_a=\frac{1}{\sqrt{2}}(|h\rangle+|v\rangle)$
and Bob prepares a single photon $b$ in the state
$|\varphi\rangle_b=\frac{1}{\sqrt{2}}(|h\rangle+|v\rangle)$. They
send their photons $a$  and $b$  into the cavities through the ports
$a_{in}$ and $b_{in}$, respectively. By choosing the outcomes with
the same parity, they can improve the fidelity of the entanglement
of the atomic ensembles shared. Let us detail the principle of our
EPP as follows.

If the four-ensemble system  $E_{A_1}E_{B_1}E_{A_2}E_{B_2}$  is in
the state $|\Psi_4\rangle_1$, Alice and Bob will obtain the outcomes
with the same parity when they measure the parity of their atomic
ensembles with our PCDs. If both Alice and Bob obtain an even-parity
outcome, the state of the four-ensemble system becomes $
|\Psi'_4\rangle_1=\frac{1}{\sqrt{2}}
(|GSGS\rangle+|SGSG\rangle)_{A_1B_1A_2B_2}$.  If both Alice and Bob
obtain an odd-parity outcome, the state of the four-ensemble system
becomes $ |\Psi''_4\rangle_1=\frac{1}{\sqrt{2}}
(|GSSG\rangle+|SGGS\rangle)_{A_1B_1A_2B_2}$. These two instances
take place with the same probability  $ f^2_0/2$.  After Alice and
Bob measure the states of their ensembles $E_{A_2}$ and $E_{B_2}$
with the basis $\{\frac{1}{\sqrt{2}}(|G\rangle \pm |S\rangle)\}$,
the entangled two-ensemble pair $E_{A_1}E_{B_1}$ in the state
$|\psi^{+}\rangle_{E_{A_1}E_{B_1}}$ is obtained with or without a
single-qubit operation on either of the ensemble $E_{A_1}$ or
$E_{B_1}$.

If the four-ensemble system  $E_{A_1}E_{B_1}E_{A_2}E_{B_2}$  is in
the state $|\Psi_4\rangle_2$ or $|\Psi_4\rangle_3$,  Alice and Bob
cannot obtain the outcomes with the  same parity. That is, when one
obtains an even-parity outcome of the PCD on his/her two atomic
ensembles, the other obtains an odd-parity outcome. Alice and Bob
will obtain the states $|\psi^{+}\rangle_{E_{A_1}E_{B_1}}$ and
$|\phi^{+}\rangle_{E_{A_1}E_{B_1}}$ with the same probability  $
f_0(1-f_0)/2$ after they perform the measurement on their ensembles
$E_{A_2}$ and $E_{B_2}$ and operate the ensembles  $E_{A_1}$ and
$E_{B_1}$ with or without a single-qubit operation.

If the system $E_{A_1}E_{B_1}E_{A_2}E_{B_2}$ is initially in state
$|\Psi_4\rangle_4$, Alice and Bob will also obtain the outcomes with
the same parity when  the PCDs are applied. If the outcomes are all
even, the state of the four-ensemble system becomes $
|\Psi'_4\rangle_4=\frac{1}{\sqrt{2}}
(|GGGG\rangle+|SSSS\rangle)_{A_1B_1A_2B_2}$.  If the outcomes are
all odd, the state of the four-ensemble system becomes $
|\Psi''_4\rangle_4=\frac{1}{\sqrt{2}}
(|GGSS\rangle+|SSGG\rangle)_{A_1B_1A_2B_2}$. These two instances
take place with the same probability  $ (1-f_0)^2/2$.  After Alice
and Bob measure the states of the ensembles $E_{A_2}$ and $E_{B_2}$
with the basis $\{\frac{1}{\sqrt{2}}(|G\rangle \pm |S\rangle)\}$,
they project $E_{A_1}E_{B_1}$ into the state
$|\phi^{+}\rangle_{E_{A_1}E_{B_1}}$  with or without a single-qubit
operation on either of the ensemble $E_{A_1}$ or $E_{B_1}$.

By keeping the instances in which Alice and Bob obtain the outcomes
with the same parity, the state of the remained two-atomic-ensemble
systems $E_{A_1}E_{B_1}$ will be project into $\rho'$,
\begin{eqnarray}    %%%%%%%%%%%%%%%%%%%%%%%%%%%%%%%%Equation 21
\rho'=\frac{1}{f_0^2+(1-f_0)^2}\left[f_0^2|\psi^+\rangle\langle\psi^+|+(1-f_0)^2|\phi^+\rangle\langle\phi^+|\right].
\end{eqnarray}
The fidelity of the remained two-atomic-ensemble systems
$F^i_{p}=\frac{f_0^2}{f_0^2+(1-f_0)^2}>f_0$ when $f_0>1/2$. By
iterating our EPP several rounds, the parties can share a subset of
two-atomic-ensemble systems in a nearly maximally entangled state.
For instance, if the initial state with the fidelity $f_0>0.7$ is
used in our EPP, Alice and Bob can obtain a subset of   systems in
the state with the fidelity $F_p>0.997$ for only two rounds.

\begin{figure}[tpb]%[tpb]                                              %Figure5 4(Color online)
\begin{center}
\includegraphics[width=5.3 cm,angle=0]{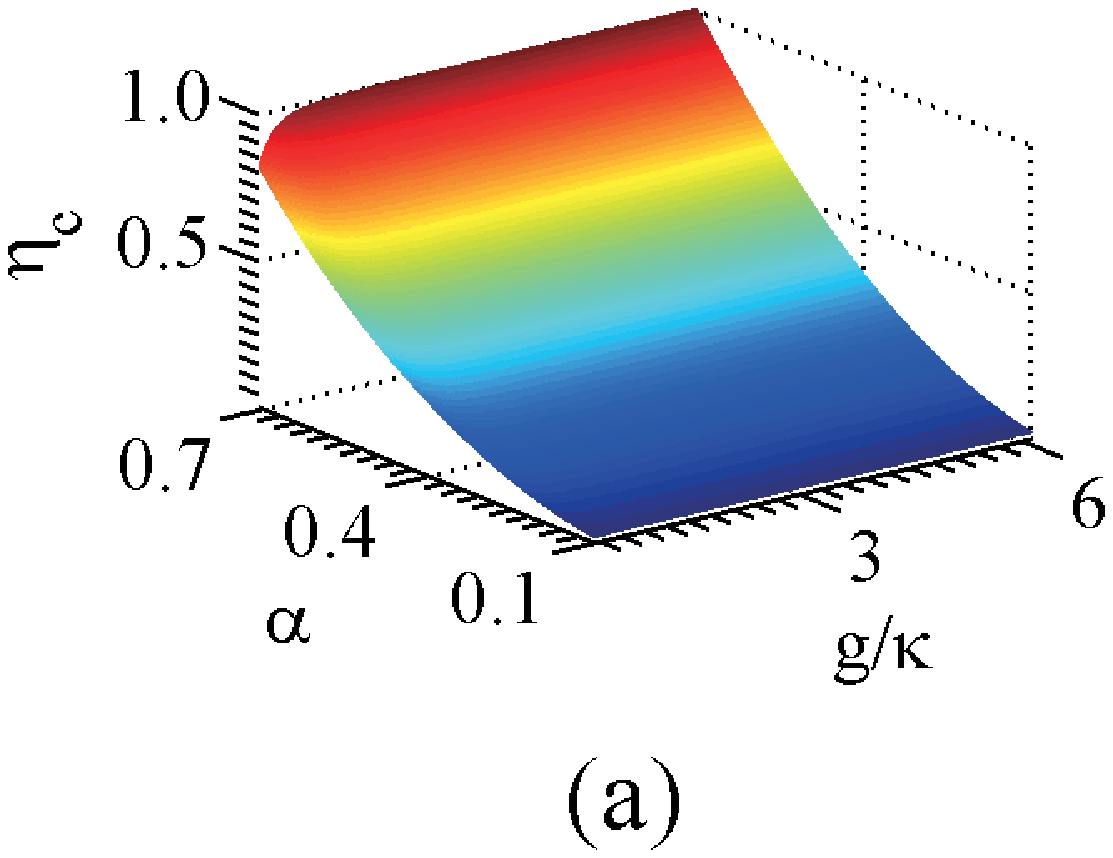}
\includegraphics[width=5.3 cm,angle=0]{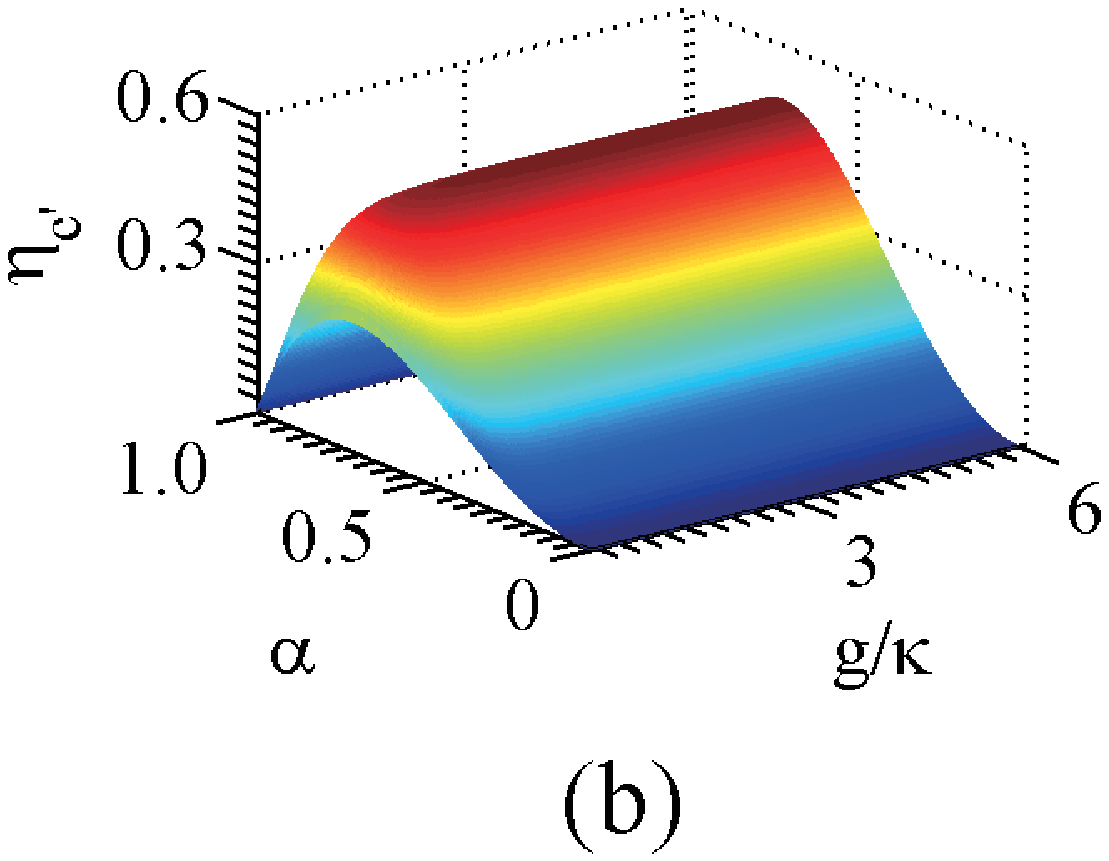}
\includegraphics[width=5.3 cm,angle=0]{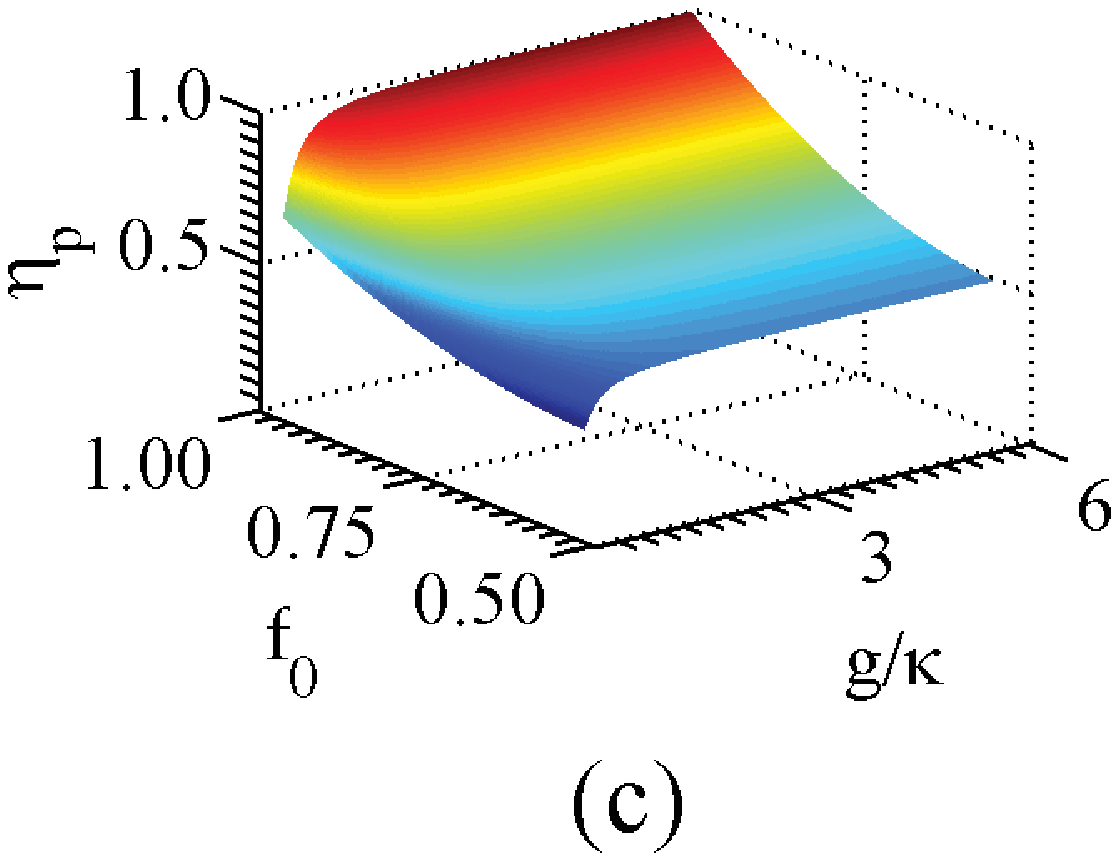}
\caption{ The efficiencies $\eta_c$, $\eta_{c'}$, and $\eta_p$ of
our optimal ECP, efficient ECP, and EPP for two-atomic-ensemble
systems, respectively. Here $|\alpha|^2+|\beta|^2=1$.}\label{fig7}
\end{center}
\end{figure}

\begin{figure}[tpb]%[tpb]                                              %Figure4 4(Color online)
\begin{center}
\includegraphics[width=5.3 cm,angle=0]{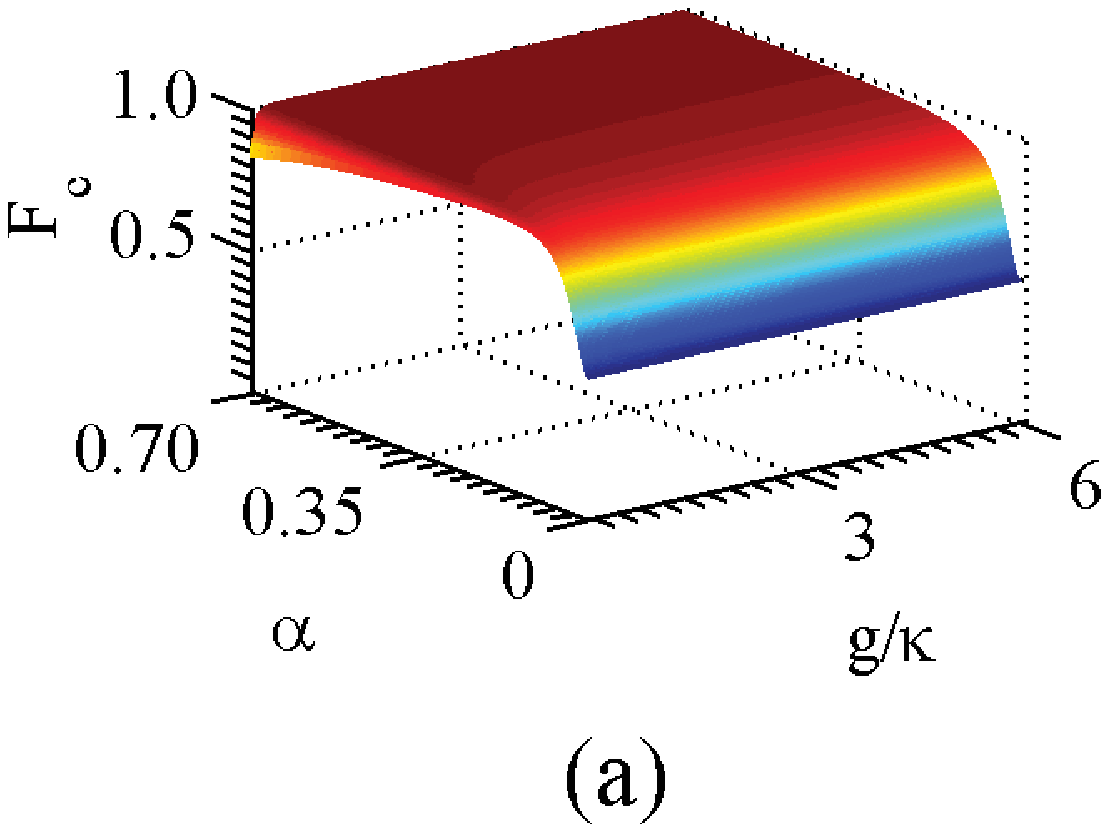}\;\;\;\;\;\;\;
\includegraphics[width=5.3 cm,angle=0]{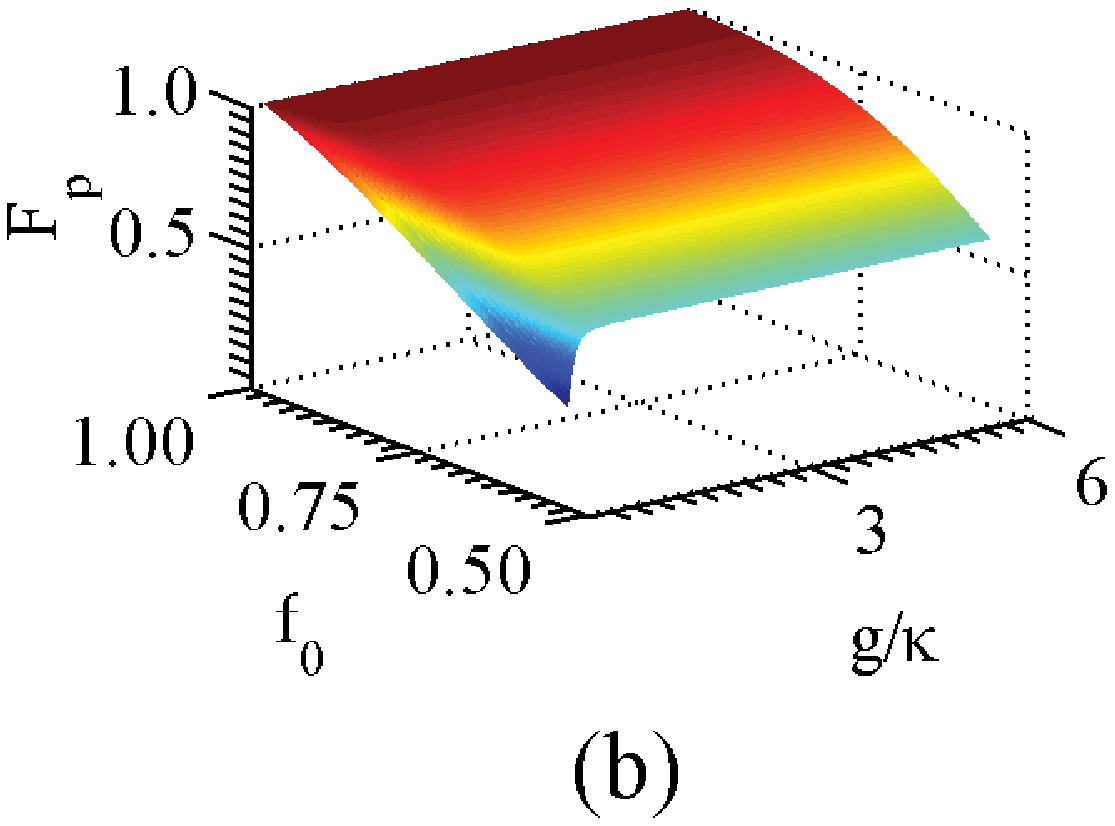}%{fig6-fp.eps}
\caption{ The fidelities of our EDPs for two-atomic-ensemble systems
as the functions of the scaled coupling strength $g/\kappa$ and the
coefficient of the initial state $\alpha$ for our optimal ECP or the
initial fidelity $f_0$ for our EPP. Here the scaled detuning
$\delta'/\kappa=0.0566$. (a) The fidelity of our optimal ECP  $F_c$.
(b) The fidelity of our EPP  $F_p$.  Here $|\alpha|^2+|\beta|^2=1$.}
\label{fig6}
\end{center}
\end{figure}

\section{ Efficiencies and fidelities of our entanglement distillation protocols}

%for two-atomic-ensemble systems

In the previous sections, our three efficient EDPs for
two-atomic-ensemble systems are proposed by using the ideal
input-output process of a single photon as a result of cavity QED.
With the optimal ECP and the efficient ECP, the parties in the
quantum communication network can get the target   systems in the
maximally entangled state. The fidelity of the systems after they
are purified with our EPP is just the same as that by the original
EPP  with the CNOT gates when only the bit-flip is involved
\cite{Bennett1996}. Then, we would like to discuss the performance
of our protocols when the practical input-output process of a photon
is considered, in which the incident photon of finite bandwidth is
used and the practical coupling strength $g$ and the decay rate
$\gamma$ of the single collective excitation state $|E\rangle$ are
taken into account.

Recently, Colombe \emph{et al}. \cite{Colombe} demonstrated the
strong atom-field coupling in an experiment in which each $^{87}$Rb
atom in Bose-Einstein condensates is identically and strongly
coupled to the cavity mode with a fibre-based cavity
\cite{Hungercavityfiber}.  In this experiment, all the atoms are
initialized to be the hyperfine Zeeman state $|5S_{1/2}, F=2,
m_f=2\rangle$. The dipole transition of $^{87}$Rb $|5S_{1/2},
F=2\rangle$ $\mapsto$ $|5P_{3/2}, F'=3\rangle$ is resonantly coupled
to the cavity mode with the maximal single-atom coupling strength
$g_0=2\pi\times215 MHz$. Meanwhile, the cavity photon decay rate is
$\kappa=2\pi\times53 MHz$ and the atomic spontaneous emission rate
of $|5P_{3/2}, F'=3\rangle$ is $\gamma_e=2\pi\times3 MHz$. The
parameters are available for our protocols, when we initialize all
the $^{87}$Rb atoms to be the state $|g\rangle\equiv|5S_{1/2}, F=1,
m_f=1\rangle$ and encode $|s\rangle\equiv|5S_{1/2}, F=2,
m_f=2\rangle$ and $|e\rangle\equiv|5p_{3/2}, F'=3, m_f=3\rangle$.
The transition between the collective states $|S\rangle$ and
$|E\rangle$ is resonantly coupled to the cavity mode, while the
transition between $|G\rangle$ and $|E\rangle$ is dipole-forbidden.
When the input-output process of a single photon is involved,  we
should use the practical reflection operator $\hat{R}(\delta')$,
instead of the ideal reflection operator $\hat{R}$ shown in Eq.
(\ref{Rideal}), to describe the reflection process of the
$|h\rangle$ polarized photon by the cavity,
\begin{eqnarray}        %%%%%%%%%%%%%%%%%%%%%%%%%%%%%%%%Equation22
\hat{R}(\delta')=|h\rangle\langle{}h|\otimes(r_0|G\rangle\langle{}G|+r|S\rangle\langle{}S|).
\label{realinputs}
\end{eqnarray}
Here $r_0$ and $r$ are the complex reflection coefficients for the
incident photon when the atomic ensemble is in the states
$|G\rangle$ and $|S\rangle$, shown in Eqs. (\ref{r}) and (\ref{r0}),
respectively. We discuss the practical fidelity and the efficiency
of our three EDPs by considering  their processes and the practical
reflection operator $\hat{R}(\delta')$ below.  In our calcuation, we
assume that linear-optical elements we used are ideal and they do
not introduce errors, similar to all the existing EPPs and ECPs
\cite{Bennett1996,Deutsch,Pan1,Simon,Pan2,shengprA2008,RenEPPLPL,
DEPPtwostep,DEPPsheng,DEPPli,DEPPdeng,DEPPShengLPL,Bennett1996ECP,Bose1999swap,Zhao2001ECP,
Yamamoto2001ECP,Sheng2008ECP,Sheng2012ECP,Ren2013ECP,RenHECP2014}.

\subsection{The practical efficiency and  the fidelity of our optimal ECP}

For our optimal ECP, only a pair of atomic ensembles in a partially
entangled pure state with known coefficients are used  and the
success of this protocol is heralded by the instance in which either
the detector $D_h$ or $D_v$ clicks. Its practical efficiency
$\eta_c$, shown in Fig. \ref{fig7} (a), is
\begin{eqnarray}     %%%%%%%%%%%%%%%%%%%%%%%%%%%%%%%%Equation 23
\eta_{c}=\frac{1}{4}\times\left(\alpha^2|r+1|^2+\frac{\alpha^4}{\beta^2}|r-1|^2+\beta^2|r_0+1|^2+\alpha^2|r_0-1|^2\right).
\end{eqnarray}
Here $\alpha$ and $\beta$ are the coefficients of the initial state
shared by the two parties Alice and Bob. To detail the influence of
the practical input-output process on the fidelity of the target
state obtained with our optimal ECP, we take the case that the
detector $D_h$ clicks as an example and get the amended fidelity
$F_c$, shown in Fig. \ref{fig6} (a), as
\begin{eqnarray}     %%%%%%%%%%%%%%%%%%%%%%%%%%%%%%%%Equation 24
F_{c}=\frac{\left|\alpha\left[r(1+\frac{\alpha}{\beta})+1-\frac{\alpha}{\beta}
\right]-\beta
\left[r_0(1+\frac{\alpha}{\beta})+1-\frac{\alpha}{\beta}\right]\right|^2}
     {\left|\alpha\left[r(1+\frac{\alpha}{\beta})+1-\frac{\alpha}{\beta}\right]\right|^2
     +\left|\beta\left[r_0(1+\frac{\alpha}{\beta})+1-\frac{\alpha}{\beta}\right]\right|^2}.
\end{eqnarray}

For the ideal input-output process, $r_0 = -1$ and $r = 1$, the
efficiency $\eta^i_{c}$ and the fidelity $F^i_{c}$ of our optimal
ECP are
\begin{eqnarray}     %%%%%%%%%%%%%%%%%%%%%%%%%%%%%%%%Equation 25
\eta^i_{c}=2|\alpha|^2,\quad\quad\quad\quad\quad F^i_{c}=1.
\end{eqnarray}

\subsection{The practical efficiency and the fidelity of our efficient ECP}

With the practical reflection operator, the fidelity of the
efficient ECP with unknown parameters is still maximal and
$F_{c'}=1$ when the two pairs of the partially entangled pure states
are identical, because the parties keep the instances in which a
$|v\rangle$ polarized photon is detected by Bob, and the detector
$D_v$ does not click if both of the two ensembles $E_{B_1}E_{B_2}$
are in the same state, e.g., $|GG\rangle$ ($|SS\rangle$). The
efficiency (success probability) of the ECP, shown in Fig.
\ref{fig7} (b), depends on the reflection coefficients and it can be
expressed as
\begin{eqnarray}     %%%%%%%%%%%%%%%%%%%%%%%%%%%%%%%%Equation 26
\eta_{c'}=\frac{(|\alpha|^2-|\alpha|^4)(|r_0-r|^2)}{2}.
\end{eqnarray}

In the ideal case, $r_0 = -1$ and $r = 1$, the efficiency
$\eta^i_{c'}$ of our optimal ECP is
\begin{eqnarray}     %%%%%%%%%%%%%%%%%%%%%%%%%%%%%%%%Equation 27
\eta^i_{c'}=2|\alpha\beta|^2=2(|\alpha|^2-|\alpha|^4).
\end{eqnarray}

\subsection{The practical efficiency and the fidelity of our EPP}

In our EPP for the nonlocal atomic ensembles $E_{A_1}E_{B_1}$ and
$E_{A_2}E_{B_2}$, the coincidental clicks of the detectors $D_h$
($D_v$ ) and $D'_h$ ($D'_v$ ), are used to denote the instances
which are kept by Alice and Bob, followed by some local operations
and measurements on the two ensembles $E_{A_2}E_{B_2}$. That is, the
success probability with the practical reflection operator  becomes
\begin{eqnarray}     %%%%%%%%%%%%%%%%%%%%%%%%%%%%%%%%Equation 28
\eta_p=f_0^2c_{00}+f_0(1-f_0)c_{01}+f_0(1-f_0)c_{10}+(1-f_0)^2c_{11},
\end{eqnarray}
with the coefficients
$c_{00}=\frac{1}{2}|r_0r|^2+\frac{1}{32}[|(r+r_0)^2|^2+|(r-r_0)^2|^2]$,
$c_{11}=\frac{1}{4}[|r_0|^4+|r|^4]+\frac{1}{32}[|(r+r_0)^2|^2+|(r-r_0)^2|^2]$,
and $c_{01}=\frac{1}{8}[|r_0(r+r_0)|^2+|r(r+r_0)|^2]=c_{10}$. When
the detectors $D_h$ and $D'_h$ click, with the practical reflection
operator, the fidelity of the target systems $E_{A_1}E_{B_1}$ is
\begin{eqnarray}     %%%%%%%%%%%%%%%%%%%%%%%%%%%%%%%%Equation 29
F_p=\frac{f_0^2c'_{00}+f_0(1-f_0)c'_{01}}{f_0^2c'_{00}+f_0(1-f_0)c'_{01}+f_0(1-f_0)c'_{10}+(1-f_0)^2c'_{11}},
\end{eqnarray}
with the coefficients $c'_{00}=2|r_0r+\frac{1}{4}(r+r_0)^2|^2$,
$c'_{11}=|r_0^2+\frac{1}{4}(r+r_0)^2|^2+|r^2+\frac{1}{4}(r+r_0)^2|^2$,
$c'_{01}=\frac{|r+r_0|^4}{2}$, and
$c'_{10}=|r_0(r+r_0)|^2+|r(r+r_0)|^2$. The  efficiency and the
fidelity of our EPP are shown in Fig. \ref{fig7}(c) and Fig.
\ref{fig6}(b), respectively.

When $r_0 = -1$ and $r = 1$, the efficiency $\eta^i_{p}$ and the
fidelity $F^i_{p}$ of our EPP are
\begin{eqnarray}     %%%%%%%%%%%%%%%%%%%%%%%%%%%%%%%%Equation 30
\eta^i_{p}=f_0^2+(1-f_0)^2,\quad\quad\quad\quad\quad
F^i_{p}=\frac{f_0^2}{f_0^2+(1-f_0)^2}.
\end{eqnarray}

\subsection{The performance of our EDPs with current experimental parameters}

The efficiencies and the fidelities of our optimal ECP, efficient
ECP, and EPP with the experimental parameters
$(\kappa,\gamma)/2\pi=(53,3.0)MHz$  \cite{Colombe} are shown in
Figs. \ref{fig7} and \ref{fig6}, respectively, as the functions of
the scaled coupling strength $g/\kappa$ and the coefficient of the
initial state $\alpha$ ($f_0$ for EPP), where the scaled detuning
$\delta'/\kappa=\gamma/\kappa=0.0566$ is used. One can see that the
performances (both the efficiencies and the fidelities) of our EDPs
become better with a larger scaled coupling strength $g/\kappa$.
Both the fidelities $F_c$ and  $F_p$ of the optimal ECP and the EPP
increase  with the  coupling strength $g$ for $g/\kappa<0.8$ and are
almost robust to the variation of $g$ for $g/\kappa>0.8$  which
leads to $g^2/\kappa\gamma>11.3$. For example, when $g/\kappa>0.4$
and $\alpha>0.2$, the efficiency $\eta_{c'}$ of our efficient ECP
$\eta_{c'}>0.065$ which is $84.3\%$ of the ideal efficiency of this
ECP $\eta^i_{c'}=2|\alpha\beta|^2$ when $\alpha=0.2$.  When
$g/\kappa>0.8$ and $\alpha>0.2$, $\eta_{c'}/\eta^i_{c'}>95.6\%$.
When $g/\kappa>0.8$ and $f_0>0.7$,
  the fidelity and the efficiency of our EPP become $F_p>0.843$ and
$\eta_{p}>0.531$ which are $99.7\% $ and $91.6\%$ of those with the
ideal input-output process, respectively. Colombe \emph{et al}.
\cite{Colombe} demonstrated the strong atom-field coupling with
$g/\kappa>4$, which means our EDPs are feasible  with current
techniques.

\section{Discussion and summary}

In our EDPs, the perfect single-sided cavity, consisting of an ideal
mirror with $100\%$ reflection and a partially reflective mirror,
might remain challenging. One can take the methods developed in
\cite{Hungercavityfiber,singleside1,singleside2} to implement an
approximately single-sided cavity. The non-zero transmission of the
ideal mirror along with the mirror scattering and absorption will
lead to the photon loss of a probability $p_{loss}$.  An additional
photon filtering mechanism of the probability $p_{loss}$ on the
$|v\rangle$ polarized photon is needed to make the input-output
process faithful \cite{filtering}, although this slightly decreases
the overall success probability. When the photon pulse duration $T$
satisfies $T\kappa\gg1$, the temporal mode of the output pulse is
basically the same as that of the input one leading to a faithful
input-output process \cite{Duan2004QIP}. In fact, with the maximal
detuning $\delta'_{max}=\gamma=0.0566\kappa$,  the fidelities and
the efficiencies of our three EDPs can achieve the values higher
than $90\%$ of those obtained with the ideal PCDs. Certainly, we
assume that the local operation on the single photon with
linear-optical elements is ideal, similar to all the existing EPPs
and ECPs
\cite{Bennett1996,Deutsch,Pan1,Simon,Pan2,shengprA2008,RenEPPLPL,
DEPPtwostep,DEPPsheng,DEPPli,DEPPdeng,DEPPShengLPL,Bennett1996ECP,Bose1999swap,Zhao2001ECP,
Yamamoto2001ECP,Sheng2008ECP,Sheng2012ECP,Ren2013ECP,RenHECP2014,reviewECP}.

Compared with the previous ECPs for atom systems
\cite{atomECP2,atomECP3,atomECP4}, our optimal ECP and efficient ECP
for  nonlocal  atomic ensembles have some advantages. Our optimal
ECP for atomic ensembles in a partially entangled pure state with
known parameters has the optimal success probability as the same as
that of the ECP for two-photon systems \cite{Ren2013ECP} and it is
achieved by the detection of a single photon which interacts with a
single-sided cavity one time. It can be used to distill the
entanglement of each pair of atomic ensembles and does not require
additional atomic ensembles, which relaxes the difficulty of its
implementation in experiment largely. Our efficient ECP can be used
to distill a subset of atomic ensembles in a  maximally entangled
state from those in a partially entangled pure state with unknown
parameters, resorting to a PCD which is constructed in a simple way
and involves only one effective input-output process of a single
photon, not two or more.  Moreover, the success of our efficient ECP
is heralded by the individual detection of one $|v\rangle$ photon in
each node, independent of the scaled coupling strength $g/\kappa$,
which is far different from  the existing cavity-involved ECPs
\cite{atomECP2,atomECP3,atomECP4}. By iteration of the concentration
process, our efficient ECP has the maximal success probability,
compared with other ECPs for the quantum systems in an
 entangled pure state with unknown parameters.

Our EPP for nonlocal atomic ensembles in a mixed entangled state is
more efficient than that in \cite{ZhaoChenQR},  where the
entanglement purification for atomic systems is completed with the
EPP for two-photon systems in \cite{Pan1} conditioned on the
effective quantum memory \cite{Qmemory2009}. In \cite{Zhao2010EPP},
a CNOT gate for the two ensembles in each node is used to perform
the EPP and it doubles the efficiency, compared with that in
\cite{ZhaoChenQR}, while the two ensembles are required to be placed
so close to each other that the CNOT gate can be performed
faithfully. This requirement is not needed when the PCDs are used to
perform our EPP and the efficiency of our EPP equals to that in
\cite{Zhao2010EPP} in the ideal case.

In actual implementations, errors can always take place. The
efficiencies and  the fidelities of our three EDPs are influenced by
some experimental factors, such as the detector's efficiency, the
decay of the radiation to non-cavity modes, the impurities of the
single-photon sources, and so on. For heralded single-photon sources
based on the probabilistic correlated photon generation with a PDC
source, we should make the average pair production levels much less
than one to avoid producing multiple pairs that lead to the impurity
in the heralded channel \cite{EisamanSingle-photon}. Besides, with
the development of the deterministic single-photon source
\cite{EisamanSingle-photon}, the impact of the impurities of the
single-photon sources can be further reduced. The photon loss due to
the cavity mirror scattering and absorption and the nonunit
efficiency of the detectors will decrease the efficiency of the
input-output process and thus will reduce the efficiency (success
probability) of our EDPs. Fortunately, all our EDPs succeed
conditioned on the detection of a single photon and the instances
with photon loss can be picked out according to the response of the
detectors.

%, which is different from others.

In summary, we have investigated the entanglement distillation for
two-atomic-ensemble systems in single-sided cavities for the first
time and  proposed three efficient entanglement distillation
protocols, including an optimal ECP for a partially entangled pure
state with known parameters, an efficient ECP for an unknown
partially entangled pure state with a PCD which is constructed in a
simple way, and an EPP for a mixed entangled state. These EDPs have
higher fidelity and efficiency with current experimental techniques,
compared with the existing EDPs for atomic systems and
atomic-ensemble systems, and they are useful for quantum
communication network and quantum repeaters.

\section*{Acknowledgments}

This work is supported by the National Natural Science Foundation of
China under Grant Nos. 11174039, 11174040, and 11474026,
NECT-11-0031, and the Open Foundation of State Key Laboratory of
Networking and Switching Technology (Beijing University of Posts and
Telecommunications) under Grant No. SKLNST-2013-1-13.

%%%%%%%%%%%%%%%%%%%%%%% References %%%%%%%%%%%%%%%%%%%%%%%%%

\end{document}